\documentclass[a4paper,aps,prl,showpacs,twocolumn,superscriptaddress]{revtex4}

\usepackage{graphicx,t1enc}
\begin{document}
\title{Promiscuity and the Evolution of Sexual Transmitted Diseases}

\author{Sebasti\'an Gon\c{c}alves}
\affiliation {Instituto de F\'{\i}sica, Universidade Federal do Rio Grande do
Sul, Caixa Postal 15051, 90501-970 Porto Alegre RS, Brazil}
\author{Marcelo Kuperman}
\affiliation{Centro At{\'o}mico Bariloche and Instituto Balseiro, 8400 S. C.
de Bariloche, Argentina}
\author{Marcelo Ferreira da Costa Gomes}
\affiliation {Instituto de F\'{\i}sica, Universidade Federal do Rio Grande do
Sul, Caixa Postal 15051, 90501-970 Porto Alegre RS, Brazil}

\date{\today}
\begin{abstract}
We study the relation between different social behaviors and the onset of
epidemics in a model for the dynamics of sexual transmitted diseases.  The model
considers the society as a system of individual sexuated agents that can be
organized in couples and interact with each other.  The different social
behaviors are incorporated assigning what we call a promiscuity value to each
individual agent. The individual promiscuity is taken from a distributions and
represents the daily probability of going out to look for a sexual partner,
abandoning its eventual mate. In terms of this parameter we find a threshold for
the epidemic which is much lower than the classical fully mixed model
prediction, i.e. $R_0$ (basic reproductive number) $= 1$. Different forms for
the distribution of the population promiscuity are considered showing that the
threshold is weakly sensitive to them. We study the homosexual and the
heterosexual case as well.
\end{abstract}

We have introduced recently a model for the evolution of a sexual transmitted
disease where the social behavior is incorporated by means of what we call the
promiscuity variable\cite{gk02}, i.e. the daily probability of an agent to dismiss
its permanent mate and go out to look for a sexual intercourse.  The dynamics is
of the SIR type: a susceptible (S) agent can become infected (I) after a sexual
intercourse with an infectious one, with probability $\beta$ (the per sexual
contact probability of get infected by an infectious agent). The infected
individuals remain infectious for a period $\tau$, when they are removed (R) by
death.  The system will model the dynamic network of social contacts, where each
element of the network is considered as a sexually active subject.  The social
behavior element is included via the promiscuity and the fraction of singles.
In principle, individuals are grouped in couples and additionally a controlled
proportion of singles, ($\rho_s$), is considered.  The variable $p_i$, called
the ``promiscuity'', is what determines an individual's tendency to dismiss
his/her mate and go out --or just to go out in the case of the singles-- to look
for an occasional intercourse; more precisely $p_i$ is the probability of trying
to meet an occasional partner on each intercourse opportunity or time step of
the simulation (which, for the sake of simplicity, we take as a day).  The
$p_i$'s are randomly taken from a chosen distribution and assigned to each
individual; we start with a semi-Gaussian distribution of width equal to $\hat
p$, but we will compare later others distributions.  Those who happened to go out
chose a partner at random and if the latter happened to go out too, the
occasional couple is made. Therefore there is no social structure {\it a
priori}.  The web of contacts is constructed dynamically during the simulation
depending on the percentage of singles and the population promiscuity.  Most of
the results are for the homosexual case, but we considered the heterosexual case
too.  The details of the simulation are given in \cite{gk02}, and the main result
of this model can be expressed in the following relation, relating the four
parameters of the model at the epidemic transition:
$$
  \hat p^2_c \beta \tau = {0.36, 0.48, 0.65}
$$
respectively for
$$
  \rho_s= {0, 0.5, 1}
$$
where $\hat p$ is a per day probability, $\beta$ is a per intercourse
probability, and $\tau$ is in days.  This relation gives us the following values
for the mean promiscuity at the critical point: 0.025, 0.029, and 0.034,
respectively for the three values of $\rho_s$, and for $\beta = 1$ and $\tau =
1yr$.

\begin{figure}[ht]
\centering \includegraphics[width=7cm, clip=true]{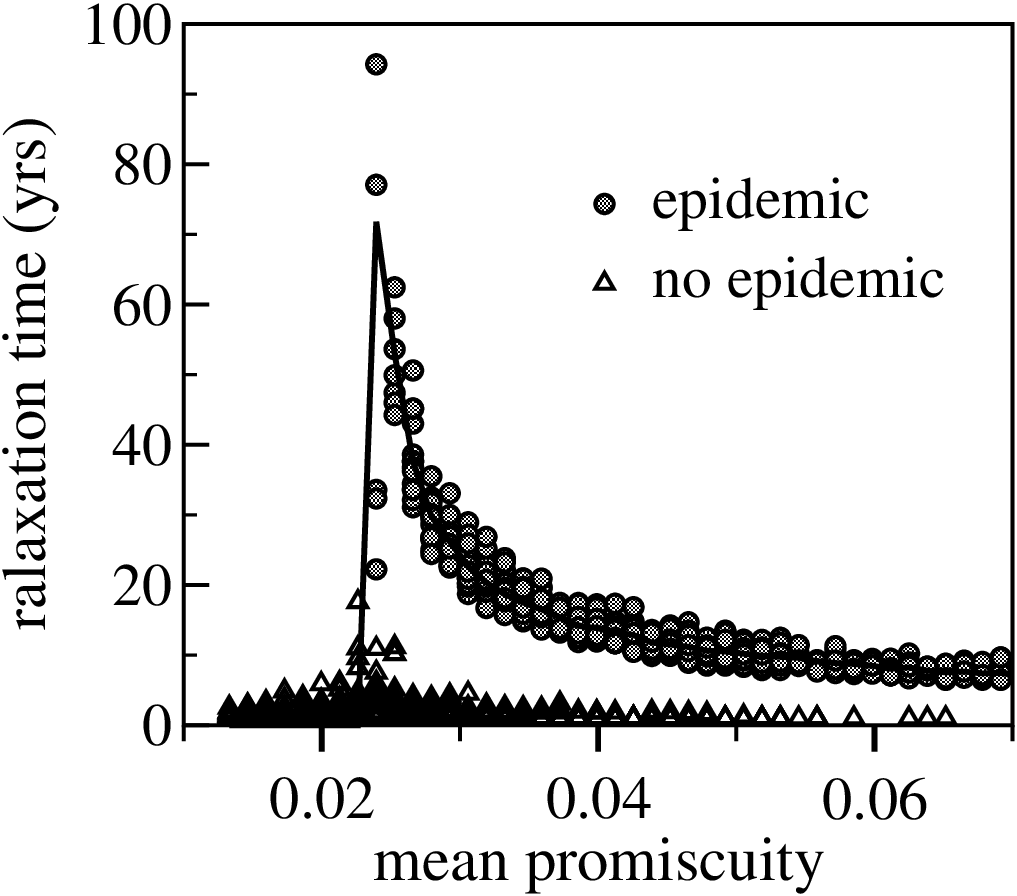}
\caption{Relaxation time as a function of average promiscuity for
$\rho_s = 0$, for $\beta = 1$ and $\tau = 1 yr$.}
\label{time}
\end{figure}

Some authors~\cite{fh63,gra83,new02} have pointed to the connection between
epidemics and percolation. In our case there is no network {\it a priori}, no
topology with fixed links between nodes (agents); instead, the sexual network is
continuously built and destroyed as time goes by, depending upon the individual
promiscuity, that are controlled by the promiscuity distribution, and the
fraction of singles.  Therefore the threshold can be understood as a dynamic
percolation.  This is supported not only by the sharp transition of the
asymptotic number of death at the critical promiscuity value, but by the
divergence of the time needed to get that asymptotic number of death (remember
that there is no other cause of death in the present model) at the same
promiscuity value, as can be seen in Fig.~\ref{time}.  All of this has a strong
resemblance with a percolation transition.

%fig 1

\begin{figure}[ht]
\centering \includegraphics[width=7cm, clip=true]{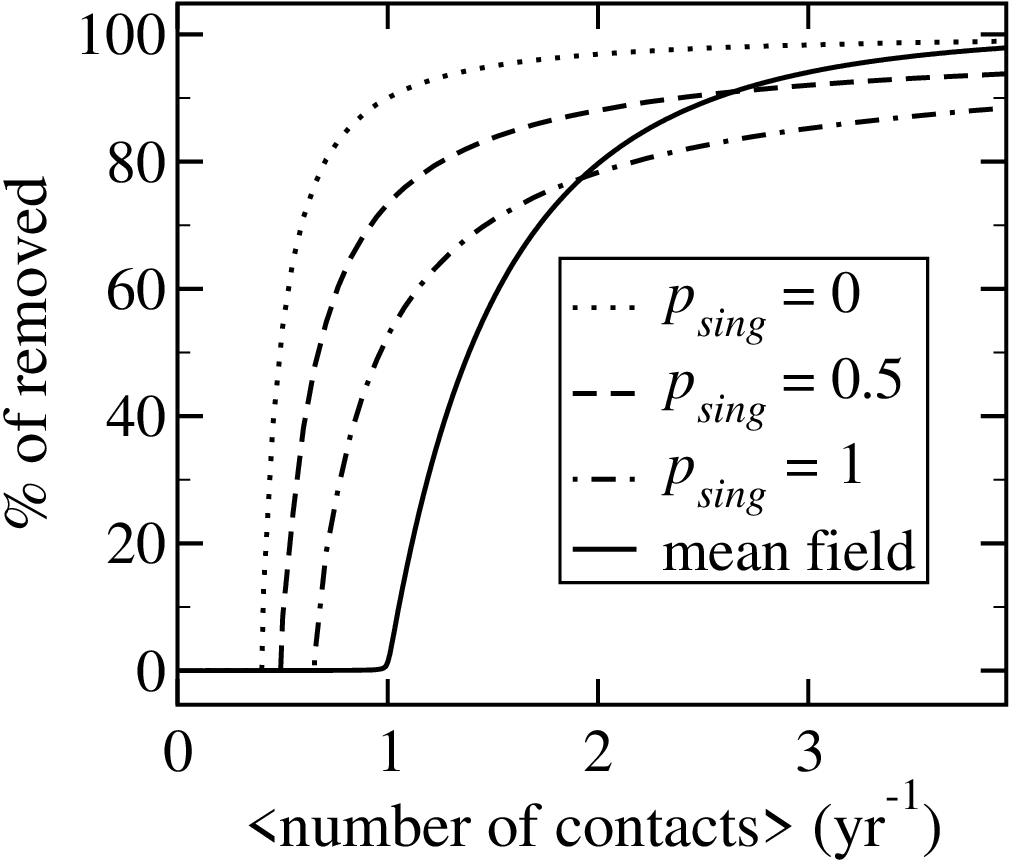}
\caption{Asymptotic percentage of removed agents as a function of the average
number of contacts per year, for three different values of $\rho_s$
Comparison with the numerical solution of the fully mixed SIR model
($\beta=1~day^{-1}$, $\tau=1~yr$). Simulations are made with 100000
subjects and 100 realizations.}
\label{pc}
\end{figure}

In the classical implementation of the SIR model
(see~\cite{Murray}), the dynamics of the susceptible, infectious, and
removed classes is governed by the following ordinary nonlinear
equations:
$$
  \frac{ds}{dt}=-r i s, \hspace{0.5cm} \frac{di}{dt} = r i s - a i,
                        \hspace{0.5cm} \frac{dr}{dt} = a i
$$
where $r$ is the infectious rate and $a$ is the removal rate per unit time; $r$
can be expressed as the product of $\beta$ and $c$, the mean number of sexual
contact per unit time, and $a$ is the inverse of the infectious time $\tau$. The
onset of the epidemic happens when the basic reproductive number $R_0 = \beta
\tau c$ is equal to 1.  Therefore it is interesting to compare the threshold and
the number of asymptotic death of the standard fully mixed model with the
present results. In the former case we have to resort on numerical integration
to solve the equations above. We do this up to a time were no new susceptible
get infected (in practice when $dS < 1/N$) and we plot the final number of
removed agents against the average number of individual sexual contact per unit
time, assuming that $\beta = 1$ and $\tau = 1yr$.  The results are plotted in
Fig~\ref{pc} together with the prediction of our model.  In order to do the
comparison we have obtained from the simulations the relation between the average
promiscuity and the average number of contacts per unit time (the latter depends
almost quadratically on the former one).

%fig 2

Comparing them we see that the threshold predicted by the numeric solution of
our model is considerably lower, possibly due to the effect of fluctuation that
the uniformly mixed SIR model does not take into account. Moreover, the
divergence between the latter and simulations increases when couples got into
the scene.  This more sophisticated version speed up the epidemic~\cite{gk02},
so we conclude that the epidemic dynamics, in general, can not be reduced to
take the average number of contacts as the only relevant parameter; as for our
results, $\rho_s$ for example, is a relevant parameter too, and all of this
reinforce the idea of using simulation as the appropriated technique to study
epidemics.

%fig 3
\begin{figure}[ht]
\centering \includegraphics[width=7cm, clip=true]{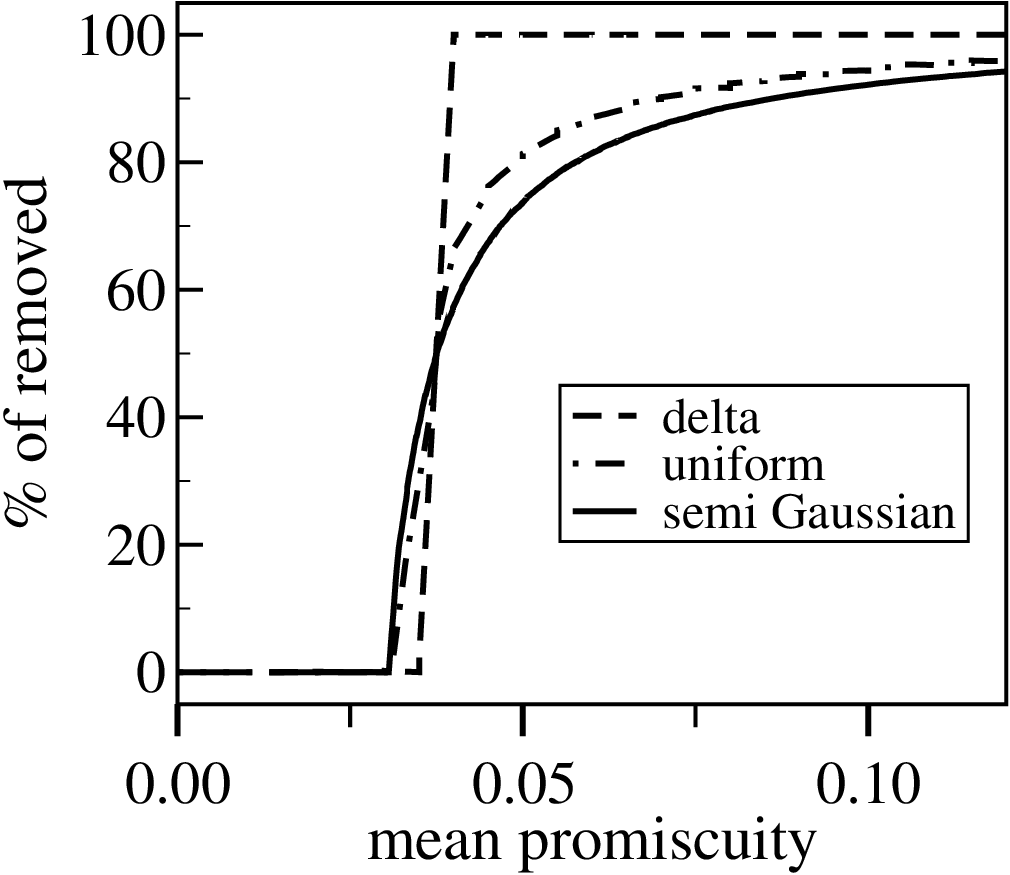}
\caption{Asymptotic percentage of removed agents as a function of the mean
promiscuity of the population $\langle p\rangle$ for several
promiscuity distributions. Parameters are: $\tau=1yr$ and
$\beta=1day^{-1}$.  Numerical results obtained with 100000 subjects,
100 realization, and for $\rho_s=1$ case (all singles).}
\label{distri}
\end{figure}

The results summarized in Fig.~\ref{time} and Fig.~\ref{pc} are for a
semi-Gaussian distribution of promiscuity, so the natural question is ``How
sensitive to the choice of distribution are the results presented here?'' In
Fig.~\ref{distri} we plot the results for a delta ($p_i =$ constant) and a
uniform distribution together with the semi-Gaussian distribution, all of them
normalized to the corresponding mean promiscuity $\langle p\rangle$.  We can see
in Fig.~\ref{distri} that the uniform and the semi-Gaussian distributions
present almost the same threshold, while the delta distribution has its
threshold slightly shifted to a larger promiscuity, due to the absence of
heterogeneity. In the other hand we have previously shown~\cite{gk02} that a
power law distribution --more appropriated to the real word~\cite{lens01}--,
produce a contrary effect: a threshold shifted to a smaller promiscuity value.
Nevertheless even in this latter case it is a weak effect, so we conclude that
irrespective of the distribution of promiscuity, that in turn governs the
distribution of contacts, the mean value is the most relevant measurements to
mark the onset of the epidemic, while other moments (especially the standard
deviation) seem not to be important, in contradiction with recent
discussions~\cite{ml01,ek02,lm01}.

A final word about the solution of the fully mixed classical implementation of
the SIR model: generally speaking it is said that the limitation of this
approach came from the fully mixed assumption and from assigning the same value
of infectivity, and the same number of per year contacts to all
individuals. However, apart from that limitations, there is one that comes
first, it cames from the method of solution that assumes the population as
continuous quantities without fluctuations. Let's see: if we take the delta
distribution, i.e. the same promiscuity for all agents, we have the model in the
condition described above: each agent takes a potential partner at random from
the whole population, so it is indeed fully mixed, and the parameters $p,
\beta$, and $\tau$ are the same for everybody.  However if we compare the curve
of the classical SIR model solution of Fig.~\ref{pc} with the results for the
delta distribution in Fig.~\ref{distri} the differences are obvious. Not only
the threshold is overestimated by the classical approximation, the simulation
predicts a sharp transition separating the no epidemic region from the epidemic
one with no survivors at all. We emphasize that, apart from the limitation of
the model, we are comparing here the exact solution of it (simulations) with an
continuous -approximated- solution of the same model (the ODE equations)

\begin{figure}
\centering \includegraphics[width=7cm, clip=true]{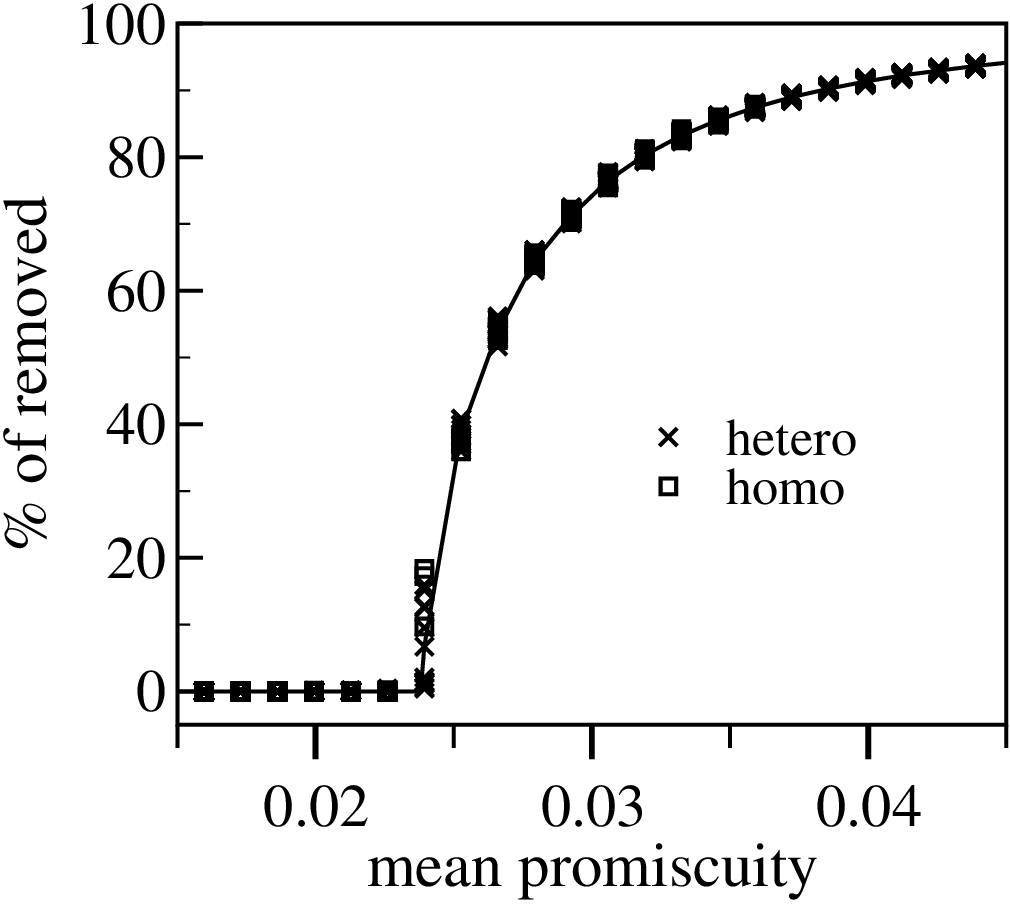}
\caption{Asymptotic percentage of removed agents as a function of mean
population promiscuity $\langle p \rangle$.  Numerical result obtained
with 100000 subject and 100 realization for $\rho_s=0$ case.
Heterosexual and homosexual cases are displayed with symbols for each
realization; the line is an average over them.}
\label{hetero}
\end{figure}

So far we have assumed an heterogeneous population in terms of the promiscuity,
but homogeneous on all of other respects, more appropriated for a gay community
for example. What happens if we consider a heterosexual population?  Taking the
all married case version, but now 50\% males and 50\% females, we have the same
output as for the homosexual case, as can be observed in Fig.~\ref{hetero}. The
point is that even when we have two subgroups in the population there is no
asymmetry (the same promiscuity distribution, infectivity and,$\tau$ for all
agents, whether they are males or females) so the results is no surprising. If
we broke the symmetry we would expect a different behavior between the hetero
and homo cases as we plan to study in the near future.

%fig 4

Summarizing we have presented some results for the spread of sexual transmitted
diseases, based on a model that takes into account the social behavior.  The
model predicts a threshold condition quite different that what the classical SIR
models predict. The onset of the epidemic is associated in the present model
with a percolation threshold of a dynamical kind, because there is no static
network of contact.  The present model gives a direct connection between disease
and social parameters and the outbreak threshold, which is quantitative
equivalent to a basic reproductive number $R_0 = 0.64$, for the singles case
(0.4 for the all married case), much lower than that of the fully mixed model
prediction ($R_0 = 1$).  On the other hand we have shown that the onset of
epidemic depends mainly on the average number of contact, but seems to be
independent of the exact form of the distribution, in opposition to recent
analysis that include the second moment in the basic reproductive number of
heterogeneous populations.  Finally the threshold is sensitive to the fraction
of singles, and this can have important consequences in modeling real data.

S.G. and M.K. acknowledge support from Coordena\c{c}\~ao de
Aperfei\c{c}oamento de Pessoal de N\'{\i}vel Superior (CAPES, Brasil)
and Secretar\'{\i}a de Ciencia, Tecnolog\'{\i}a e Innovaci\'on
Productiva (SETCIP, Argentina).

\end{document}